\documentclass[reprint,amssymb,amsmath,aps,showpacs,showkeys,letter,prb]{revtex4}

\begin{document}

\title{Theoretical method for the study of the excited states of a system}


\author{Ram\'on Alain Miranda Quintana}
\email{alain@instec.cu}
\affiliation{Department of Radiochemistry, Higher Institute of Technologies and Applied Sciences, Ave. Salvador Allende y Luaces, Quinta de los Molinos, Plaza de la Revoluci\'on, POB 6163, 10600, Havana, Cuba.}

\begin{abstract}
A novel, exact, theoretical method for the study of the excited states of a system is presented. It is demonstrated how to transform the excited state problem of one Hamiltonian into the ground state problem of an auxiliary one.  From this, a new exact density functional suitable for excited states is constructed. These results make the excited states of a system accessible through all ground state theoretical techniques.
\end{abstract}


\maketitle

In spite of the development in the course of the years of many sophisticated and powerful theoretical tools, the variational principle of Rayleigh and Ritz (RR) keeps its central role in quantum theory. This principle, besides its many practical applications, has also served as starting point for the development of new theoretical methods, the density functional theory (DFT) being an example of this \cite{hohemb,levy,parr}. But the classical formulation of the variational principle is restricted to the study of ground states or to the first excited states having a symmetry which is different from the symmetry of the ground state. Given the importance of the analysis of the excited states of a system, a great effort has been put in the search for generalizations of the RR principle in ways that can be used to study them \cite{mcdonald,theophi,kohn1,kohn2,kohn3}. Here we demonstrate a set of theorems that allow us to apply the original RR principle to excited states. In doing so, from a Hamiltonian $H_{0}$ and with some knowledge of its ground state, a new Hamiltonian $H_{1}$ is built, which has the characteristic that its ground state exactly coincides with the first excited state of $H_{0}$. A DFT type variational principle for the new Hamiltonian is obtained, which enable us to construct a new exact density functional suitable for the study of excited states. These results make the excited states of a system accessible through all ground state techniques.

Let us start giving some definitions. Let $H_{0}$ be a Hamiltonian and $\mathbb{E}_{0}$ the set of its eigenvalues. We will denote by $\mathbb{V}_{0}$ the set of the eigenvectors of $H_{0}$, if $E_{k}\in\mathbb{E}_{0}$,$\mathbb{V}_{0}^{E_{k}}$ will represent the subset of $\mathbb{V}_{0}$ associated with the eigenvalue $E_{k}$. The set $\{|E_{k}\beta\rangle\}$ corresponds to an orthogonal basis of $\mathbb{V}_{0}^{E_{k}}$ where $\beta$ can be interpreted as the eigenvalues of a set of observables which, along with the energy, constitute a complete set of observables for the system under study \cite{dirac}. Let us suppose that $E_{0}=min \mathbb{E}_{0}$ and that we know a $\{|E_{0}\beta\rangle\}$. In this case we can build a Hamiltonian $H_{1}$ according to:
\begin{equation}\label{eqn_1}
    H_{1} = H_{0}+K\sum_{\beta}|E_{0}\beta\rangle\langle E_{0}\beta|
\end{equation}
In this expression $K$ is a real constant that will be analyzed later. In full analogy as it has been done before, the sets $\mathbb{E}_{1}$, $\mathbb{V}_{1}$ and $\mathbb{V}_{1}^{E_{k}}$  corresponding to $H_{1}$ can be defined.

Now, starting from the definition of $H_{1}$, some of its properties will be analyzed. In the following, Halmos'$\blacksquare$ indicates the conclusion of a proof.

\textbf{Theorem 1:} If $E_{1}=min(\mathbb{E}_{0}\backslash \{E_{0}\})$, then: $\forall\psi$, $\langle\psi|H_{1}|\psi\rangle\geq min\{E_{1},E_{0}+K\}$

\textbf{Proof:} Because $H_{0}$ is an observable we can write \cite{dirac}:
\begin{equation}\label{eqn_2}
    \forall|\psi\rangle, |\psi\rangle=\sum_{E_{k}\beta}c(E_{k}\beta)|E_{k}\beta\rangle;E_{k}\in\mathbb{E}_{0},c(E_{k}\beta)=\langle E_{k}\beta|\psi\rangle
\end{equation}
Where $\sum_{E_{k}\beta}|c(E_{k}\beta)|^{2} =1$.

Then:
\begin{equation}\label{eqn_3}
    \langle\psi|H_{1}|\psi\rangle = \langle\psi|H_{0}|\psi\rangle+K\sum_{\beta}|c(E_{0}\beta)|^{2}
\end{equation}

Doing $\sum_{\beta}|c(E_{0}\beta)|^{2} \equiv t$ we will have that:
\begin{equation}\label{eqn_4}
    \forall|\psi\rangle, \langle\psi|H_{1}|\psi\rangle \geq min_{\mathbb{T}_{\psi}} \langle\varphi|H_{0}|\varphi\rangle+Kt
\end{equation}
The minimum is searched over the set $\mathbb{T}_{\psi}=\{|\varphi\rangle:\sum_{\beta}|\langle E_{0}\beta|\varphi\rangle|^{2} =t\}$.

It is then evident that, provided $\langle\varphi|\varphi\rangle = 1$: $min_{\mathbb{T}_{\psi}} \langle\varphi|H_{0}|\varphi\rangle = E_{0}t+E_{1}(1-t)$ so:
\begin{equation}\label{eqn_5}
\forall|\psi\rangle, \langle\psi|H_{1}|\psi\rangle \geq \epsilon(t) = E_{1}+(E_{0}+K-E_{1})t
\end{equation}
Every linear functional $\epsilon(t)$ will reach its lower value in the frontier of its domain of definition, in this case, in $t=0$ or in $t=1$. Checking that, $\epsilon(0)=E_{1}$ and $\epsilon(1)=E_{0}+K$ ends up proving the Theorem. $\blacksquare$

\textbf{Theorem 2:} $\mathbb{V}_{0}\subseteq\mathbb{V}_{1}$

\textbf{Proof:} $|\psi\rangle\in\mathbb{V}_{0}\Rightarrow|\psi\rangle= \sum_{\beta}c(E_{k}\beta)|E_{k}\beta\rangle;E_{k}\in\mathbb{E}_{0},c(E_{k}\beta)=\langle E_{k}\beta|\psi\rangle$ So:

\begin{subequations}
\label{eqn_6}
\begin{eqnarray}
H_{1}|\psi\rangle&=&H_{0}|\psi\rangle+K\sum_{\beta'}|E_{0}\beta'\rangle\langle E_{0}\beta'|\psi\rangle  \label{eqn_6a}\\
H_{1}|\psi\rangle&=&H_{0}|\psi\rangle+K\sum_{\beta'}|E_{0}\beta'\rangle\sum_{\beta}c(E_{k}\beta)\langle E_{0}\beta'|E_{k}\beta\rangle \label{eqn_6b}
\end{eqnarray}
\end{subequations}
Where $\langle E_{0}\beta'|E_{k}\beta\rangle=\delta_{E_{0}E_{k}}\delta_{\beta'\beta}$, being $\delta_{ij}$ the Kronecker's delta.

Let us consider the cases: a)$E_{k} \neq E_{0}$ and b)$E_{k} = E_{0}$.\\

a)$E_{k} \neq E_{0}\Rightarrow \langle E_{0}\beta'|E_{k}\beta\rangle = 0 \quad\forall \beta, \beta'$, then:
\begin{equation}\label{eqn_7}
H_{1}|\psi\rangle=H_{0}|\psi\rangle=E_{k}|\psi\rangle\Rightarrow|\psi\rangle\in\mathbb{V}_{1}
\end{equation}

Thus, an eigenket of $H_{0}$ corresponding to an eigenvalue $E_{k} \neq E_{0}$ will be as well an eigenket of $H_{1}$ corresponding to the same eigenvalue.

b)$E_{k} = E_{0}\Rightarrow \langle E_{0}\beta'|E_{k}\beta\rangle = \langle E_{0}\beta'|E_{0}\beta\rangle = \delta_{\beta'\beta}$, then:
\begin{equation}\label{eqn_8}
    H_{1}|\psi\rangle=H_{0}|\psi\rangle + K\sum_{\beta'}c(E_{0}\beta')|E_{0}\beta'\rangle = E_{0}|\psi\rangle + K|\psi\rangle = (E_{0}+K)|\psi\rangle\Rightarrow|\psi\rangle\in\mathbb{V}_{1}
\end{equation}

Thus, an eigenket of $H_{0}$ corresponding to the eigenvalue $E_{0}$ will be as well an eigenket of $H_{1}$ corresponding to the eigenvalue $E_{0}+K$. $\blacksquare$

From this proof it follows that $(\mathbb{E}_{0}\backslash\{E_{0}\})\cup\{E_{0} + K\}\subseteq \mathbb{E}_{1}$.

\textbf{Theorem 3:} $K>0\Rightarrow(\mathbb{E}_{0}\backslash\{E_{0}\})\cup\{E_{0} + K\} = \mathbb{E}_{1}$

\textbf{Proof:} (by \emph{reductio ad absurdum}) Let us suppose that:
\begin{equation}\label{eqn_9}
    \exists \tilde{E}\notin(\mathbb{E}_{0}\backslash\{E_{0}\})\cup\{E_{0} + K\}: \exists |\tilde{E}\rangle\neq|0\rangle;H_{1}|\tilde{E}\rangle=\tilde{E}|\tilde{E}\rangle
\end{equation}

From Theorem 1 and the RR principle: $K>0\Rightarrow\tilde{E}\neq E_{0}$.

For being $H_{0}$ an observable we can write \cite{dirac}:
\begin{equation}\label{eqn_10}
    |\tilde{E}\rangle=\sum_{E_{k}\beta}c(E_{k}\beta)|E_{k}\beta\rangle;E_{k}\in\mathbb{E}_{0},c(E_{k}\beta)=\langle E_{k}\beta|\tilde{E}\rangle
\end{equation}

But, according to Theorem 2: $\forall E_{k}\beta, E_{k}\in \mathbb{E}_{0}, |E_{k}\beta\rangle\in\mathbb{V}_{1}$ then we will have, by hypothesis:
\begin{equation}\label{eqn_11}
    \forall E_{k}\beta, E_{k}\in \mathbb{E}_{0},c(E_{k}\beta)=\langle E_{k}\beta|\tilde{E}\rangle=0
\end{equation}
The contradiction ends up proving the Theorem. $\blacksquare$

The previous results allow us to conclude with the following:

\textbf{Corollary:} $K>0\Rightarrow\mathbb{V}_{1}=\mathbb{V}_{0}$, even more, $\mathbb{V}_{1}^{\tilde{E}_{k}}=\mathbb{V}_{0}^{E_{k}}$, $\tilde{E}_{k}=E_{k}+\delta_{0k}K$

\textbf{Proof:}
From the proof of Theorem 2 it follows that $\mathbb{V}_{0}^{E_{k}}\subseteq\mathbb{V}_{1}^{\tilde{E}_{k}}$, so it suffices to proof that $\mathbb{V}_{1}^{\tilde{E}_{k}} \subseteq \mathbb{V}_{0}^{E_{k}}$.

Be $|\psi\rangle\in\mathbb{V}_{1}^{\tilde{E}_{k}}$, according to Theorem 3 this implies that $\tilde{E}_{k}\in(\mathbb{E}_{0}\backslash\{E_{0}\})\cup\{E_{0} + K\}$.

For being $H_{0}$ an observable we can write \cite{dirac}:
\begin{equation}\label{eqn_12}
    |\psi\rangle=\sum_{E_{k'}\beta}c(E_{k'}\beta)|E_{k'}\beta\rangle=\sum_{\beta}c(E_{k}\beta)|E_{k}\beta\rangle + \sum_{(E_{k'}\neq E_{k})\beta}c(E_{k'}\beta)|E_{k'}\beta\rangle; E_{k'}\in \mathbb{E}_{0}, c(E_{k'}\beta)= \langle E_{k'}\beta|\psi\rangle
\end{equation}
In this expression when $|\psi\rangle\in\mathbb{V}_{1}^{\tilde{E}_{0}}$ we will take $k=0$

From Theorem 2: $\sum_{\beta}c(E_{k}\beta)|E_{k}\beta\rangle\in\mathbb{V}_{1}^{\tilde{E}_{k}}$, then:
\begin{equation}\label{eqn_13}
    \sum_{(E_{k'}\neq E_{k})\beta}c(E_{k'}\beta)|E_{k'}\beta\rangle = |\psi\rangle - \sum_{\beta}c(E_{k}\beta)|E_{k}\beta\rangle \in\mathbb{V}_{1}^{\tilde{E}_{k}}
\end{equation}
for being $\mathbb{V}_{1}^{\tilde{E}_{k}}$ a linear subspace.

Denoting the set of all pure states of the system by $\mathbb{W}$ we will have, for being $H_{1}$ an observable \cite{dirac,suhubi}:
\begin{equation}\label{eqn_14}
    \mathbb{W} = \oplus_{k} \mathbb{V}_{1}^{\tilde{E}_{k}}
\end{equation}
where $\oplus_{k}$ stands for the direct sum of the different subspaces. But from (\ref{eqn_14}) \cite{suhubi}:
\begin{equation}\label{eqn_15}
    \mathbb{V}_{1}^{\tilde{E}_{k}}\cap\sum_{k'\neq k}\mathbb{V}_{1}^{\tilde{E}_{k'}}=\{|0\rangle\}
\end{equation}
Then:
\begin{equation}\label{eqn_16}
    |\psi\rangle=\sum_{\beta}c(E_{k}\beta)|E_{k}\beta\rangle\in\mathbb{V}_{0}^{E_{k}}\quad\blacksquare
\end{equation}
Let us analyze the case:
\begin{equation}\label{eqn_17}
    K > E_{1} - E_{0}
\end{equation}
from which we can directly infer that $K>0$.

In virtue of the properties proven before results that, with this selection of $K$, we have built a Hamiltonian whose ground state satisfies $E_{1}=min\mathbb{E}_{1}$ at the same time that $\mathbb{V}_{1}^{E_{1}}=\mathbb{V}_{0}^{E_{1}}$. That is, the study of the first excited state of $H_{0}$ reduces to the study of the ground state of $H_{1}$, for which we can make use of the techniques classically prescribed for the study of minimum energy states.

Among the requirements needed for the construction of $H_{1}$, the necessity of having solutions for the ground state of $H_{0}$ that forms an orthogonal basis for this subspace stands out. This, besides been computationally expensive, can reminds us the known fact that the RR principle can be applied to the study of excited states under certain conditions \cite{levine}. In effect, if we work with the expression $min_{\mathbb{P}_{0}}\langle\psi|H_{0}|\psi\rangle $, where it has been pointed out that the minimum must be searched over the set $\mathbb{P}_{0}=\{|\psi\rangle:\langle\psi|\varphi\rangle=0 \forall|\varphi\rangle\in\mathbb{V}_{0}^{E_{0}}\}$, we will obtain the energy and eingenfunctions corresponding to the first excited state of $H_{0}$. This procedure is different from the one presented here: in the expression of the RR principle corresponding to $H_{1}$ all \textit{N} particle functions fulfilling the Pauli principle are included. That is, the expression $min\langle\psi|H_{1}|\psi\rangle$ can be used directly, exactly as the original RR principle $min\langle\psi|H_{0}|\psi\rangle$, because we are dealing with a ground state. In this new method there is no need to know the set $\mathbb{P}_{0}$.

The other condition for $H_{1}$ having the desired properties, namely $K>E_{1}-E_{0}$, is of the outermost importance. Even when $K>0$, but with $K$ not large enough to fulfill the above condition, the application of the RR principle to $H_{1}$ will end in a ground state with energy $E_{0}+K$ whose eigenfunctions will be those of the ground state of $H_{0}$. In order to ensure that (\ref{eqn_17}) holds true we must work with $K$ large enough. As a way to check if we have made a right choice is enough to determine the $min\langle\psi|H_{1}|\psi\rangle$, if this value is different from $E_{0}+K$ the chosen $K$ value is correct.

Another advantage of being able to apply the RR principle to $H_{1}$ lies in the possibility of formulate a DFT type variational principle for the first excited state of the Hamiltonian if we take, in the spirit of Levy's constrained search \cite{levy,parr}:
\begin{equation}\label{eqn_18}
    E_{1}=min_{\rho}E[\rho]
\end{equation}
Where the minimum is searched over all the \textit{N}-representable densities, being:
\begin{subequations}
\label{eqn_19}
\begin{eqnarray}
   E[\rho] &=& F_{H_{0}}[\rho]+\int\rho(r)\upsilon(r)dr \label{eqn_19a}\\
   F_{H_{0}}[\rho] &=& inf_{\psi\rightarrow\rho}\langle\psi|\left(T+W+K\sum_{\beta}|E_{0}\beta\rangle\langle E_{0}\beta|\right)|\psi\rangle \label{eqn_19b}
\end{eqnarray}
\end{subequations}
In the above expressions $\upsilon(r)$ stands for the external potential present in $H_{0}$, $T$ and $W$ correspond to the kinetic energy and particle-particle interaction operators respectively. The subindex in $F_{H_{0}}$ emphasizes the non-universality of this functional, an unfavorable feature that this method shares with another DFT extensions to excited states \cite{parr,valone,gross,lieb}. However, it must be noticed that the non-universal character is respect to the excited state density. The functional is in fact an universal bifunctional, depending on the electronic density of the excited and ground states \cite{levynagyprl,nagylevypra}.

Applications of this scheme within DFT can be expected in two different ways. First, the derivation of formal constraints (of the type of coordinate scaling, asymptotic behavior, etc.) for density functionals designed for excited states. For this, besides using traditional tools as the virial and Hellmann-Feynman theorems, we can use a novel property of this functional. The fact that $\forall K:K>E_{1}-E_{0}$ $E[\rho]$ always reach the same minimum value in the same density functions will contribute to shed some light in the formal structure of these functionals. Second, practical applications in DFT can be derived by means of the optimized potential method \cite{nagylevypra}. Here it is not a problem that the energy partition for the excited state functionals not needs to be the same that for the ground state functionals, since the adiabatic connection can help us to develop new exchange and correlation functionals suitable for excited states \cite{levygorlingpra,levynagyprl}.

To conclude let us point out that the method presented here can be extended to the study of an arbitrary excited state if we have enough information about the stationary states below it. That is, the study of $\mathbb{V}_{0}^{E_{k}}$, $E_{k}\in\mathbb{E}_{0}$ is possible using techniques designed to ground states if $\{|E_{k'}\beta\rangle\}\forall E_{k'}\in\mathbb{E}_{0},E_{k'}<E_{k}$ are known.

\begin{acknowledgments}
I am indebted to Alberto Beswick, Ernesto Altshuler, Marco Mart\'inez and Alfo J. Batista for their critical reading of the manuscript.
\end{acknowledgments}

\end{document}